\documentclass{article}
\usepackage[margin=2.5cm]{geometry}

\usepackage{rotating}
\usepackage{graphicx}
\usepackage{hyperref}

\begin{document}

\title{\bf LHCb PDF measurements}

\author{Mika Vesterinen, on behalf of the LHCb collaboration\\ Physikalisches Institut Heidelberg \\ \normalsize Proceedings of the Third Annual Large Hadron Collider Physics Conference LHCP15,\\ \normalsize Aug. 31-Sept 5, 2015 St. Petersburg, Russia.}

\maketitle

\begin{abstract}
The LHCb experiment covers a unique region of acceptance at forward rapidities in the
high energy proton-proton collisions of the LHC.
This means that measurements of particle production in LHCb are directly sensitive
to the parton distribution functions at low Bj\"orken-$x$ values.
Several measurements of inclusive $W$ and $Z/\gamma^{\ast}$ production 
with the Run-I dataset are reported in these proceedings.
Further measurements of $W$ and $Z/\gamma^{\ast}$ production in association
with inclusive jets and $b$- and $c$-tagged jets are also reported.
\end{abstract}

\section{Introduction}

The search for new physics at the LHC and future hadron colliders is 
reliant on a precise understanding of the partonic structure of the proton,
which is encoded in the parton distribution functions (PDFs).
They are constrained by a mixture of hadron collider, fixed target
and $ep$ collider data on processes for which the perturbative partonic cross sections can
be calculated to a high degree of precision.
The LHC experiments themselves play a crucial role in constraining the PDFs.
The LHCb experiment~\cite{LHCb-TDR-009} covers a unique region of kinematic acceptance~\cite{Thorne:2008am},
having full tracking, calorimeter and particle identification capabilities
in the pseudorapidity region $2 < \eta < 5$.
In hadron-hadron collisions at a centre of mass energy of $\sqrt{s}$, 
the production of a state of mass $M$ with rapidity $y$ is initiated
by partons of momentum fractions,
\begin{equation}
x_{1,2} = \frac{M}{\sqrt{s}}e^{\pm y}.
\end{equation}
LHCb measurements of vector boson production 
are sensitive down to $x \sim 10^{-5}$.
Several such measurements are reported in these proceedings.
These are based on 1~fb$^{-1}$ recorded at $\sqrt{s}=7$~TeV in 2011,
and 2~fb$^{-1}$ recorded at $\sqrt{s}=8$~TeV in 2012.
In addition to its unique coverage of forward pseudorapidities, LHCb has the most precise
luminosity determination at a hadron collider experiment.
Using a combination of beam-gas imaging and van der Meer scans,
the luminosities of the 7 and 8~TeV datasets are determined with relative
uncertainties of 1.7\% and 1.12\%, respectively~\cite{Aaij:2014ida}.

\section{Inclusive \boldmath{$W$} and \boldmath{$Z/\gamma^{\ast}$} cross section at \boldmath{$\sqrt{s}=7$}~TeV}

\begin{figure}[tb]\centering
    \includegraphics[width=0.70\linewidth]{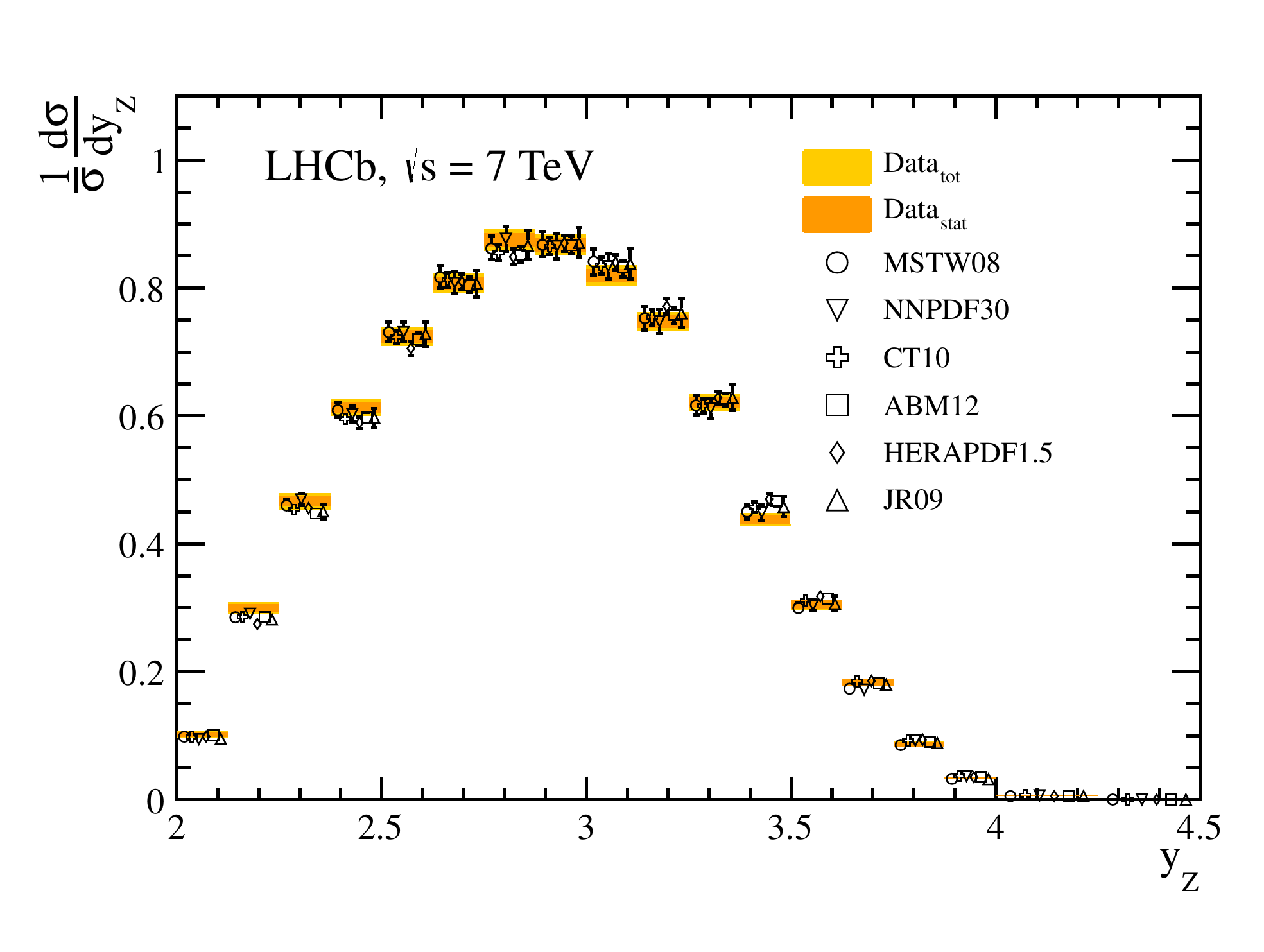}
  \caption{The $Z/\gamma^{\ast} \to \mu^+\mu^-$ cross section as a function of rapidity~\cite{LHCb-PAPER-2015-001}.}
  \label{fig:ZRap}
\end{figure}

In~\cite{LHCb-PAPER-2015-001}, LHCb reports on a measurement of the cross section
for inclusive $Z/\gamma^{\ast} \to \mu^+\mu^-$ production.
The muons must have transverse momenta in excess of 20 GeV$/c$,
and be reconstructed in the region $2 < \eta < 4.5$.
Candidates are considered within a dimuon invariant mass range between 60 and 120 GeV$/c^2$.
At least one of the muons must be matched to a single muon line at all stages
of the trigger.
Roughly 60k signal candidates are obtained, with less than 1\% background contamination.
The signal yields are corrected for the muon trigger, reconstruction and selection efficiencies,
which are measured using $Z/\gamma^{\ast} \to \mu^+\mu^-$ candidates with special requirements.
For example, the muon identification efficiencies are measured with a sample in which 
these requirements are only imposed on one of the muons.
Integrated over the kinematic acceptance defined by the above requirements,
the following cross section is obtained,
\[ \sigma(pp \to Z/\gamma^{\ast} \to \mu^+\mu^-) = 76.0 \pm 0.3_{\rm stat} \pm 0.5_{\rm syst} \pm 1.0_{\rm beam} \pm 1.3_{\rm lumi}~\mathrm{pb}, \]
where the third uncertainty relates to the knowledge of the LHC collision energy.
The cross section is also measured as a function of the transverse momentum,
$\phi_{\eta}^{\ast}$~\cite{Banfi:2010cf}, and rapidity of the dimuon pair.
The latter is shown in Figure~\ref{fig:ZRap}, 
in comparison to predictions from the \texttt{FEWZ} NNLO generator~\cite{Li:2012wna,Gavin:2010az}
with the ABM12~\cite{Alekhin:2013nda}, CT10~\cite{Lai:2010vv}, HERA1.5~\cite{Aaron:2009aa}, JR09~\cite{JimenezDelgado:2008hf}, MSTW08~\cite{Martin:2009iq} and NNPDF3.0~\cite{Ball:2014uwa} PDF sets.
LHCb has also measured $Z/\gamma^{\ast} \to e^+e^-$ production at $\sqrt{s} = 7$~TeV~\cite{Aaij:2012mda} and 8~TeV~\cite{Aaij:2015vua}, and $Z/\gamma^{\ast} \to \tau^+\tau^-$ production at $\sqrt{s} = 7$~TeV~\cite{Aaij:2012bi}.

\begin{figure}[tb]\centering
  \includegraphics[width=0.70\linewidth]{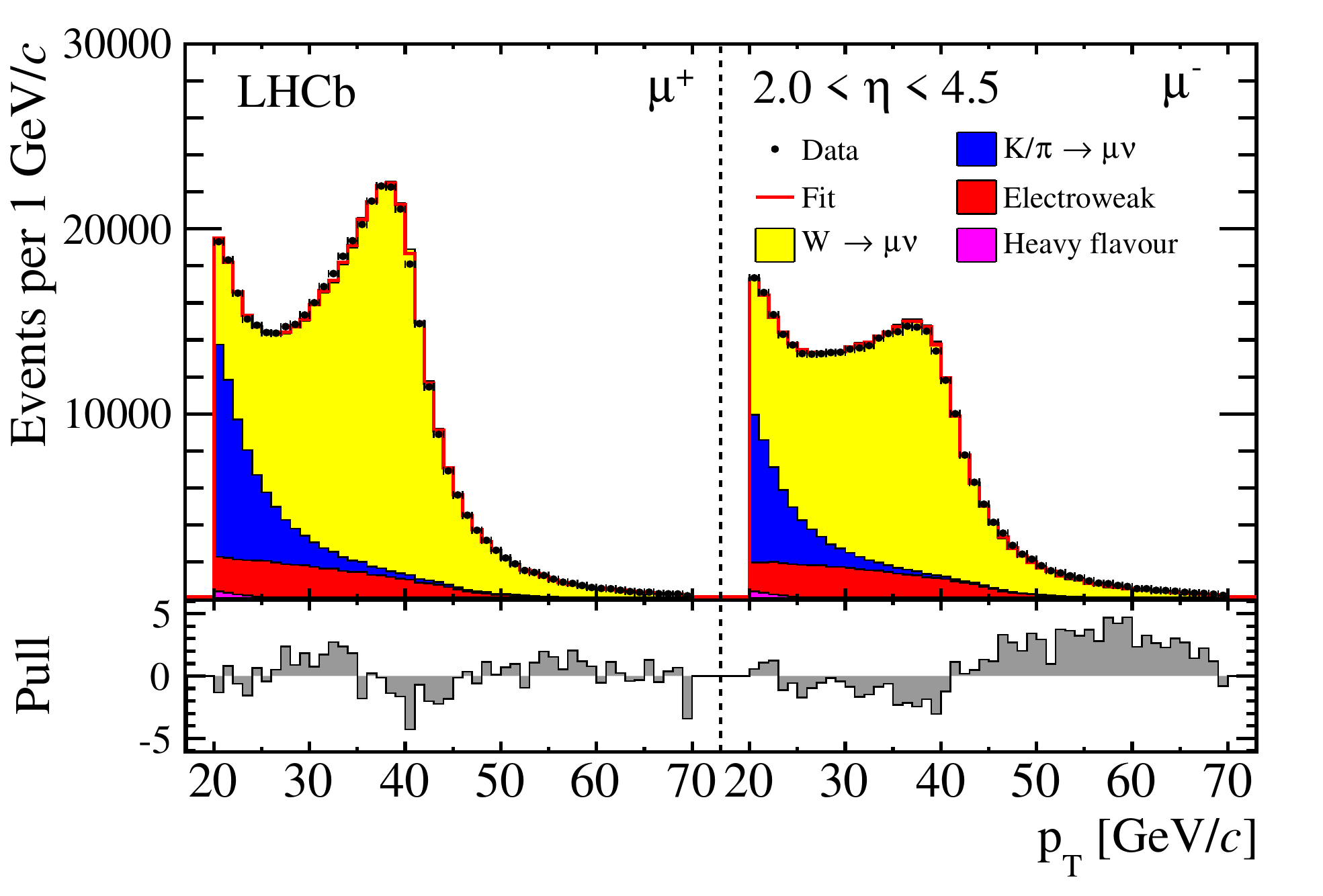}
    
  \caption{The muon $p_T$ spectrum in the (left) $W^+ \to \mu^+\nu$ and (right) $W^- \to \mu^-\bar{\nu}$ candidates~\cite{LHCb-PAPER-2014-033}.}
  \label{fig:muPT}
\end{figure}

\begin{figure}[tb]\centering
  \includegraphics[width=0.70\linewidth]{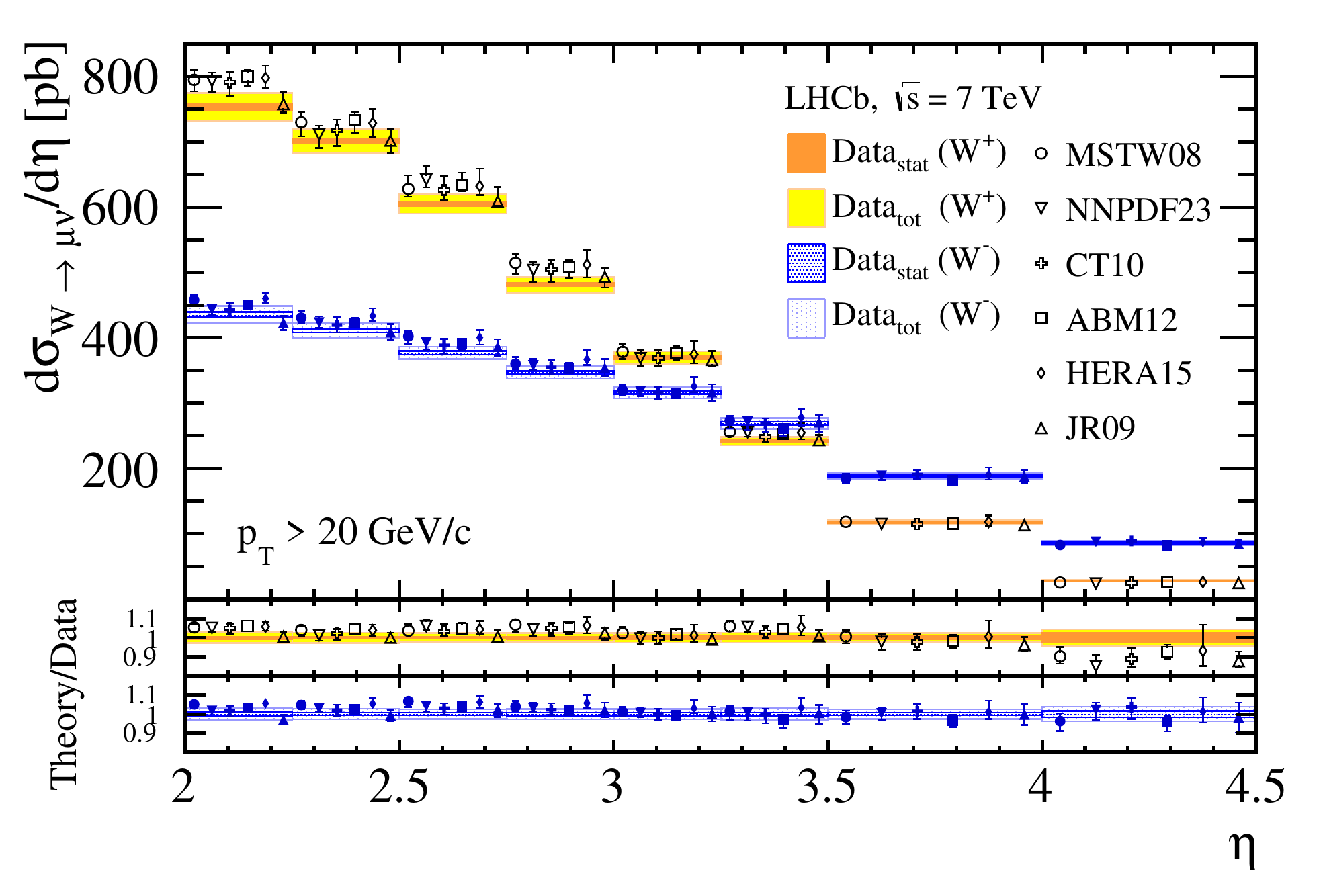}
    
  \caption{The $W^+ \to \mu^+\nu$ and $W^- \to \mu^-\bar{\nu}$ cross sections as a function of the muon $\eta$~\cite{LHCb-PAPER-2014-033}.}
  \label{fig:muSigma}
\end{figure}

LHCb reports a measurement of $W$ production at $\sqrt{s}=7$~TeV~\cite{LHCb-PAPER-2014-033}.
The kinematic requirements on the muon are the same as those that are applied to the muons
in the $Z/\gamma^{\ast}$ study.
Further isolation requirements are needed to control the level of background from 
in-flight decays of hadrons to muons.
Candidates are vetoed if another high $p_T$ muon is present in the event, in order to suppress the background from $Z/\gamma^{\ast} \to \mu^+\mu^-$.
The signal yields are extracted by fitting the muon $p_T$ spectra
as shown in Figure~\ref{fig:muPT}. The signal purity is around 70\%.
The signal yields are corrected for all sources of inefficiency using the same methods
described above in the context of the $Z/\gamma^{\ast} \to \mu^+\mu^-$ study.
Integrated over the kinematic acceptance defined above, the following cross sections are measured,
\[ \sigma(pp \to W^+  \to \mu^+\nu ) = 861.0 \pm 2.0_{\rm stat} \pm 11.2_{\rm syst} \pm 14.7_{\rm lumi}~\mathrm{pb},\]
\[ \sigma(pp \to W^-  \to \mu^-\bar{\nu} ) = 675.8 \pm 1.9_{\rm stat} \pm 8.8_{\rm syst} \pm 11.6_{\rm lumi}~\mathrm{pb},\]
and the following ratio is obtained,
\[ \frac{\sigma(pp \to W^+  \to \mu^+\nu )}{\sigma(pp \to W^-  \to \mu^-\bar{\nu} )} = 1.274 \pm 0.005_{\rm stat} \pm 0.009_{\rm syst}.\]
Figure~\ref{fig:muSigma} shows the two separate cross sections as a function of the muon $\eta$, compared to predictions with the same six PDF sets described above.
Four different cross section ratios of $W^+$, $W^-$ and $Z/\gamma^{\ast}$ 
are reported~\cite{LHCb-PAPER-2015-001}.  
These are determined with higher experimental precision due to cancelling systematic uncertainties, notably in the luminosity and in the muon reconstruction efficiencies. 
These ratios will help to constrain the flavour structure of the proton, in particular the strange quark content and symmetry.

\section{Low mass Drell-Yan production.}

\begin{figure}[tb]\centering
    \includegraphics[width=0.70\linewidth]{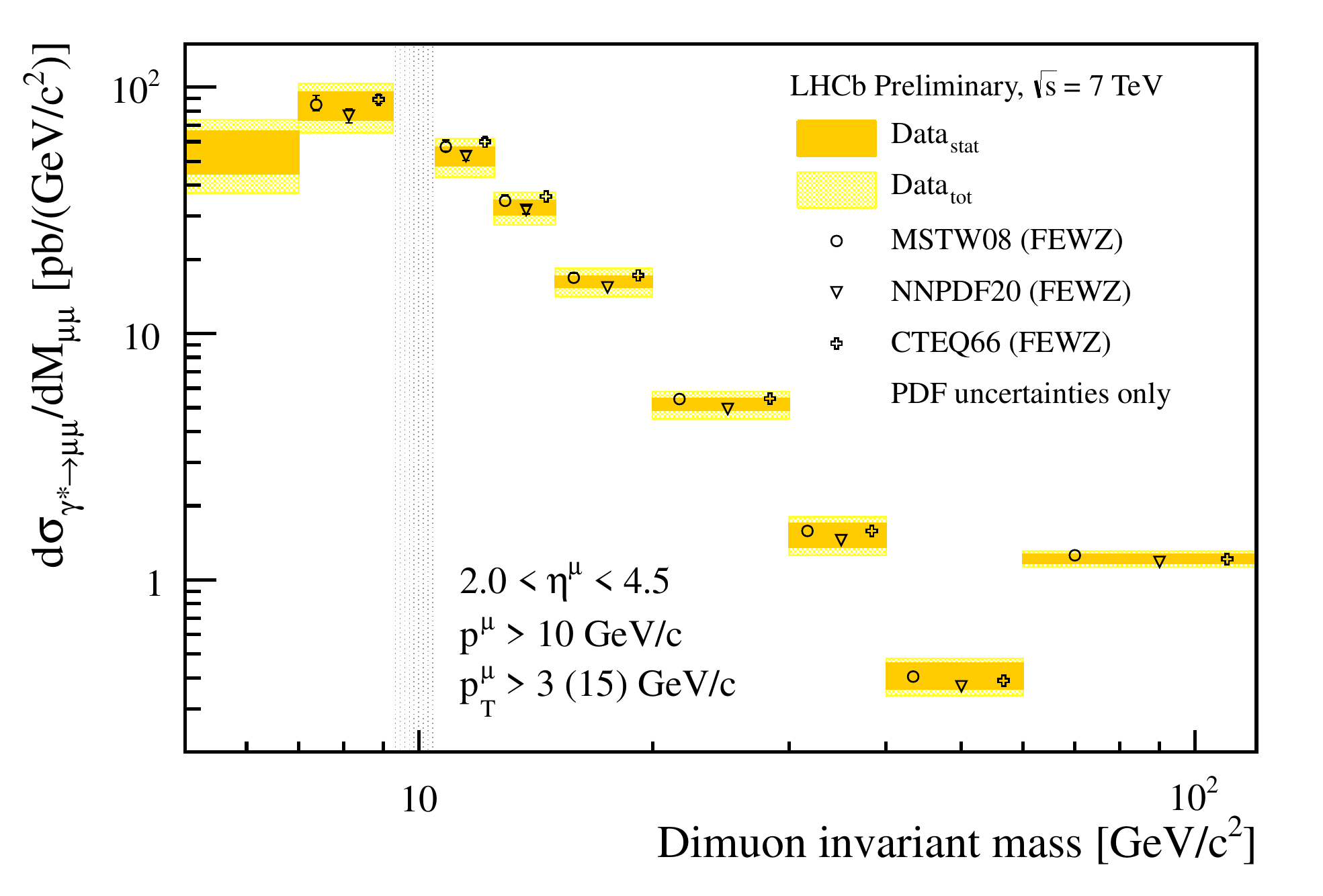}
    
  \caption{
    The $Z/\gamma^*$ cross section as a function of the dimuon invariant 
mass~\cite{LHCb-CONF-2012-013}}
  \label{fig:LowMassSigma}
\end{figure}

In~\cite{LHCb-CONF-2012-013}, LHCb reported a measurement
of $\gamma^* \to \mu^+\mu^-$ production at $\sqrt{s}=7$~TeV, and covering invariant masses as low
as 5~GeV$/c^2$, which corresponds to $x$ values below $10^{-5}$.
This measurement is based on a $37$~pb$^{-1}$ recorded during the 2010 run.
The signal yield is extracted by fitting the muon isolation distributions to subtract the hadronic backgrounds.
Figure~\ref{fig:LowMassSigma} shows the measured cross section as a function of the dimuon invariant mass,
which is in good agreement with the QCD calculations.

%

\clearpage
\section{Vector boson production in association with jets}

\begin{figure}[tb]\centering
    \includegraphics[width=0.80\linewidth]{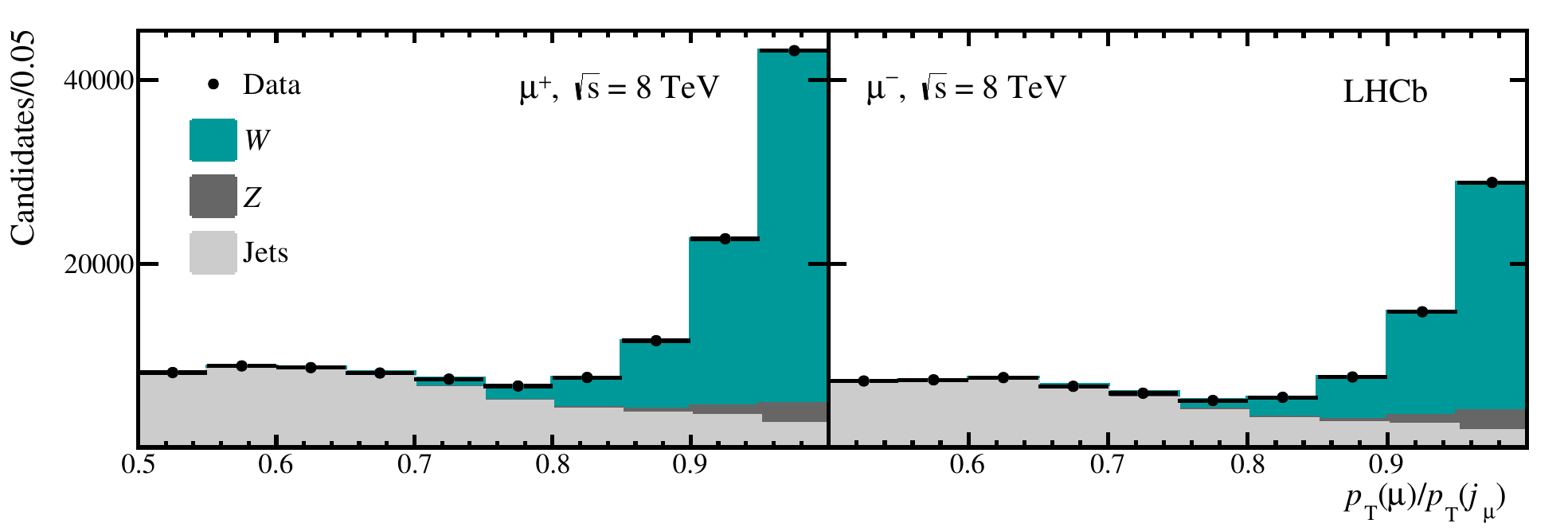}
    
  \caption{
    The $p_T(\mu)/p_T(\mathrm{jet},\mu)$ distribution of (left) $W^+$+jet and (right) $W^-$+jet
candidates, in the 2012 dataset, corresponding to a centre of mass energy of $8$~TeV.~\cite{LHCB-PAPER-2015-021}}
  \label{fig:Wjet}
\end{figure}

\begin{figure}[tb]\centering
    \includegraphics[width=0.80\linewidth]{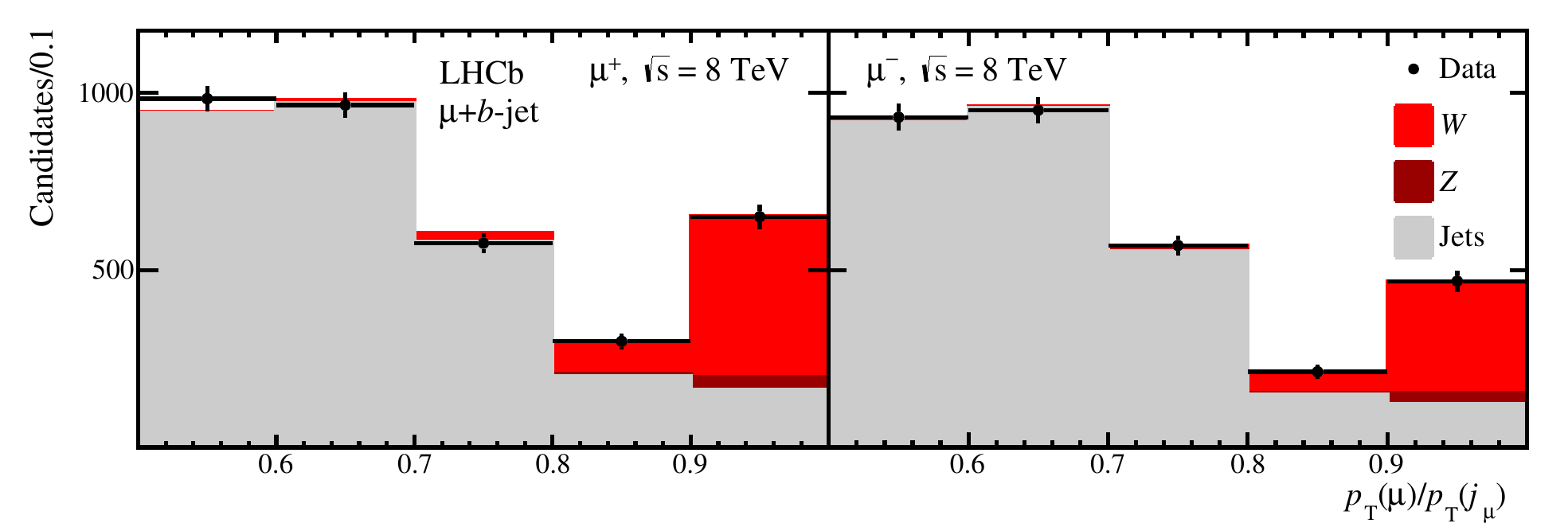}
    
  \caption{
    The $p_T(\mu)/p_T(\mathrm{jet},\mu)$ distribution of (left) $W^+$~+~$b$-jet and (right) $W^-$~+~$b$-jet
candidates, in the 2012 dataset, corresponding to a centre of mass energy of $8$~TeV.~\cite{LHCB-PAPER-2015-021}}
  \label{fig:Wb}
\end{figure}

\begin{figure}[tb]\centering
    \includegraphics[width=0.80\linewidth]{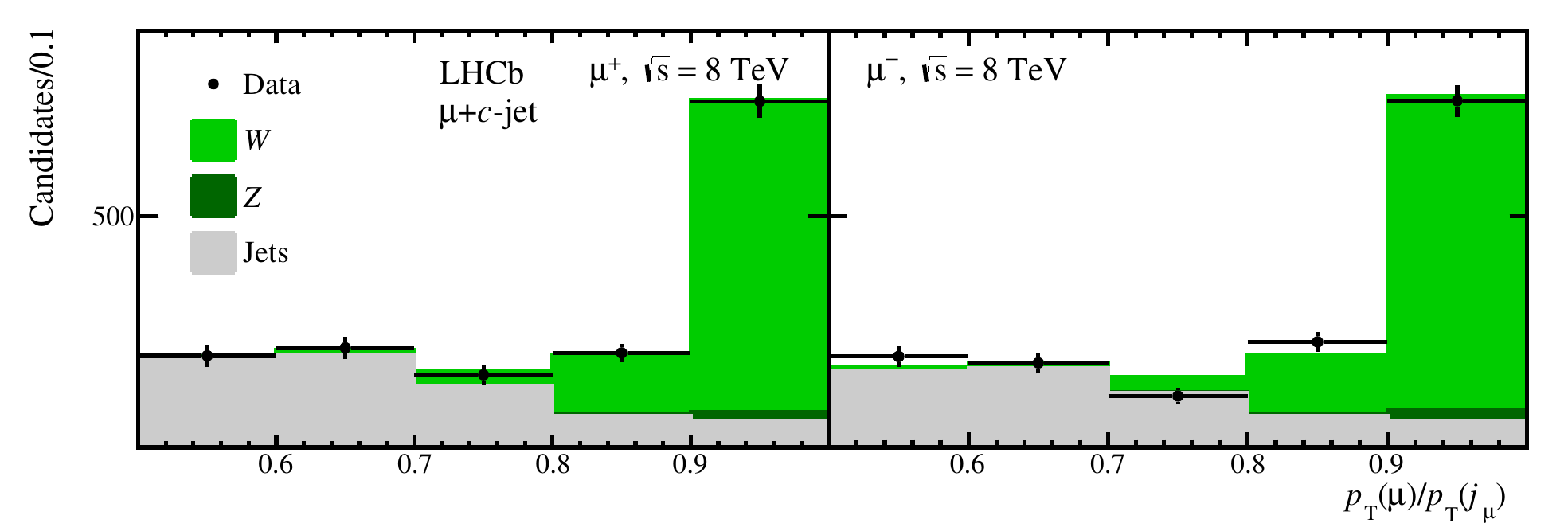}
    
  \caption{
    The $p_T(\mu)/p_T(\mathrm{jet},\mu)$ distribution of (left) $W^+$~+~$c$-jet and (right) $W^-$~+~$c$-jet
candidates, in the 2012 dataset, corresponding to a centre of mass energy of $8$~TeV.~\cite{LHCB-PAPER-2015-021}}
  \label{fig:Wc}
\end{figure}

In~\cite{LHCb-PAPER-2013-058} LHCb reports a measurement of $Z/\gamma^{\ast}$ production
in association with jets.
The selection requirements on the $Z/\gamma^{\ast} \to \mu^+\mu^-$ candidates are the same as those
applied in the inclusive measurement described above.
The jets are reconstructed with the anti-$k_T$ algorithm~\cite{Cacciari:2008gp} with a cone size of $0.5$,
as implemented in the {\sc FASTJET} package~\cite{Cacciari:2005hq}.
In~\cite{Aaij:2014gta} this measurement was extended to study $Z/\gamma^{\ast}$~+~$b$-jet production
by searching for a secondary vertex within the jet.
LHCb recently developed dedicated $b$- and $c$-jet identification algorithms~\cite{Aaij:2015yqa}.
A boosted decision tree is trained to distinguish $b$-jets from light jets,
while another is trained to distinguish between $b$- and $c$-jets.
For jets with $p_T > 20$~GeV$/c$ and $2.2 < \eta < 4.2$, it is possible to 
identify $b$-jets with an efficiency of around 65\% for a mis-id (from light jets)
rate of 1\%. For $c$-jets the corresponding efficiency with the same fake rate is around 25\%.
In~\cite{LHCB-PAPER-2015-021} LHCb reported a measurement of $W$+jet production with the full Run-I dataset.
The signal component is extracted with the use of an isolation variable that considers
the $p_T$ of the reconstructed jet which contains the muon.
Figure~\ref{fig:Wjet} shows, for the 2012 part of the dataset,
the $p_T(\mu)/p_T(\mathrm{jet},\mu)$ distribution to which a fit is performed
to determine the signal yield.
Using the identification algorithms described above, the $W$+$b$ and $W$+c components are extracted.
The $p_T(\mu)/p_T(\mathrm{jet},\mu)$ distributions are shown for $b$- and $c$-jet enriched
regions in Figures~\ref{fig:Wb} and~\ref{fig:Wc}, respectively.
All of these measurements are in agreement with QCD calculations.
The $W$+jet measurements will constrain the valence $d$-quark PDF,
while the $W$+charm measurements will constrain the strange quark PDFs,
and the $W$+beauty measurements will constrain the gluon PDF.

\section{Conclusions}
The study of particle production within the LHCb experiment probes the proton structure
in a unique region of $x$ and $Q^2$.
Compared to the other LHC experiments, LHCb is the only one that has full tracking, calorimeter
and particle identification over the pseudorapidity range $2 < \eta < 5$.
Several measurements of inclusive $W$ and $Z/\gamma^*$ production are reported.
These are complemented by further measurements of associated production of $W$ and $Z/\gamma^*$
with inclusive jets, and with $b$- and $c$-tagged jets.
These measurements have helped to constrain and reduce uncertainties on current PDF sets, while new measurements from LHCb will help to improve the precision of future PDF sets even further.

\nocite{*}
\bibliographystyle{unsrt}
\bibliography{sample}

\end{document}